\documentclass[a4]{article}
 \input{epsf}
\usepackage{lscape}
\usepackage{rotating}
\usepackage{amsmath}
\usepackage{epsfig}
\usepackage{amssymb,amsfonts}
\usepackage[all]{xy}
\usepackage{graphicx}

\numberwithin{equation}{section}

\begin{document}

\title{Massless particles on supergroups \\ and $AdS_3 \times S^3$ supergravity}

 \author{
 Jan Troost
}
 
\maketitle

\begin{center}
 \emph{Laboratoire de Physique Th\'eorique}\footnote{Unit\'e Mixte du CNRS et
    de l'Ecole Normale Sup\'erieure associ\'ee \`a l'universit\'e Pierre et
    Marie Curie 6, UMR
    8549. 
} \\
\emph{ Ecole Normale Sup\'erieure  \\
24 rue Lhomond \\ F--75231 Paris Cedex 05 \\ France}
\end{center}

\begin{abstract}
  Firstly, we study the state space of a massless particle on a
  supergroup with a reparameterization invariant action. After gauge
  fixing the reparameterization invariance, we compute the physical
  state space through the BRST cohomology and show that the quadratic
  Casimir Hamiltonian becomes diagonalizable in cohomology. We
  illustrate the general mechanism in detail in the example of a
  supergroup target $GL(1|1)$. The space of physical states remains an
  indecomposable infinite dimensional representation of the space-time
  supersymmetry algebra. Secondly, we show how the full string BRST
  cohomology in the particle limit of string theory on $AdS_3 \times
  S^3$ renders the quadratic Casimir diagonalizable, and reduces the
  Hilbert space to finite dimensional representations of the
  space-time supersymmetry algebra (after analytic
continuation). Our analysis provides an
  efficient way to calculate the Kaluza-Klein spectrum for
  supergravity on $AdS_3 \times S^3$. It may also lead to
  the identification of an interesting and simpler subsector of
  logarithmic supergroup conformal field theories, relevant to string
  theory.

 \end{abstract}

\section{Introduction}
It is believed that the holographic nature of quantum gravity
\cite{'tHooft:1993gx}\cite{Susskind:1994vu} renders anti-de Sitter
compactifications of string theory equivalent to dual gauge theories
\cite{Maldacena:1997re}. Many pairs of dual theories have been
proposed.  They often involve anti-de Sitter backgrounds of string
theory with Ramond-Ramond flux which arise in the near-brane limit of
backreacted D-branes \cite{Polchinski:1995mt}.  That makes it
desirable to compute the spectrum of string theory on these
Ramond-Ramond backgrounds. The light-cone gauge is the most efficient
gauge choice to determine these spectra at present.  Nevertheless it
remains interesting to further our understanding of the calculation of
the spectrum in conformal gauge, for instance in a Berkovits
formulation of the worldsheet string action in these backgrounds (see
e.g. \cite{Berkovits:2000fe} and references thereto). These worldsheet
conformal field theories involve supergroup or supercoset target
spaces in many interesting examples. The whole of target space
(super)symmetry is manifestly realized in these models.

Two-dimensional conformal field theories with supergroup or coset
targets are also interesting in their own right. They have been
studied from various perspectives (see e.g.
\cite{Read:2001pz,Schomerus:2005bf,Saleur:2006tf,Gotz:2006qp,
Quella:2007hr,Ashok:2009xx,Benichou:2010rk,Babichenko:2009dk,Creutzig:2010ne,Benichou:2010ts}).
One crucial feature of these theories is that they are 
logarithmic. The scaling operator is not diagonalizable on the state
space. Moreover, this feature already manifests itself in the
one-dimensional limit of these models.  Indeed, the Laplacian on a
supergroup is typically not diagonalizable on the space of
quadratically integrable functions
\cite{Huffmann:1994ah,Schomerus:2005bf,Saleur:2006tf,Quella:2007hr}. Moreover,
the space of functions is typically an infinite dimensional
indecomposable representation of the supersymmetry algebra. 
Although space-time superisometries are manifest, their representation
is intricate.

String theory in particular Ramond-Ramond backgrounds and in conformal
gauge will be built using such a conformal field theory, but it will only
make use of a physical state space determined by a BRST cohomology. It
is insensitive to BRST exact features of the worldsheet conformal
field theory. For complicated target spaces though, it can be hard to
discern what the BRST exact data in the worldsheet conformal field
theory are that one may wish to ignore.  To gain insight into this
question, we study simpler models that exhibit some of the same
crucial features.

Concretely, in this paper we compute the BRST cohomology for a
reparameterization invariant particle living on a supergroup manifold,
and investigate to what extent the curious features of the space of
quadratically integrable functions survive in the physical state
space. In section \ref{particle} we show that implementing
reparameterization invariance on the physical state space is enough to
render the quadratic Casimir diagonalizable.  We illustrate the
details of the structure of the representation space in the case of
the supergroup $GL(1|1)$ in section \ref{gl11}.  In section
\ref{psu22} we compute the full string BRST cohomology for
compactification independent states in $AdS_3 \times S^3$ string
theory with Ramond-Ramond and Neveu-Schwarz-Neveu-Schwarz flux, and
show that due to the more refined cohomology, the space of physical
states decomposes into finite dimensional representation spaces of the
supersymmetry algebra. As a byproduct, we show that this gives an
efficient derivation of the equivalence of this subsector of string theory to
supergravity, as well as a brief and manifestly supersymmetric
derivation of the Kaluza-Klein spectrum. Finally, we draw general
lessons for applications of logarithmic conformal field theories to
string theory.

\section{A massless particle on a supergroup}
\label{particle}
In this section we study a massless particle on a supergroup $G$ and argue that its
Hamiltonian becomes diagonalizable in cohomology.
\subsection{The action}
Our model for a massless particle on a supergroup $G$ 
is defined in terms of a reparameterization invariant
action.   Due to the fermionic directions in target
space, the model will be non-unitary.
The action is:
\begin{eqnarray}
S &=& \int_L d \tau  
e^{-1} \langle g^{-1} \partial_\tau g, g^{-1} \partial_\tau \rangle,
\label{action}
\end{eqnarray}
where the map $g : L \rightarrow G : \tau \mapsto g(\tau)$ maps
the worldline $L$ of the massless particle into the group manifold $G$, and $\langle .,. \rangle$ denotes
an invariant metric on a Lie super algebra ${\mathfrak g}$. We choose it
to be proportional to the supertrace in a matrix representation
of the algebra. The field $e$ is an einbein on the worldline of the particle.
After gauge fixing the reparameterization invariance through the gauge choice
$e=1$, we find the gauge fixed action:
\begin{eqnarray}
S &=& \int d \tau  
 \langle g^{-1} \partial_\tau g, g^{-1} \partial_\tau \rangle
 + \int d \tau b \partial_\tau c,
\end{eqnarray}
where we introduced the $(b,c)$ ghosts to take into account the measure
factor arising from gauge fixing.
 The physical state space in the quantum theory will be 
determined by the cohomology of the
BRST operator:
\begin{eqnarray}
Q_B &=&  c \, C_2 
\end{eqnarray}
where the quadratic Casimir $C_2$ equals the
Hamiltonian of the system. 
\subsection{The quadratic Casimir in cohomology}
\label{diagonalizable}
When we consider a particle on a supergroup, its wave-function will
correspond to a function on the supergroup.  Functions on a supergroup
can be expanded in the fermionic coordinates on which they depend. We
will study a space of functions such that the coefficients in the
fermionic coordinate expansion are quadratically integrable. We refer
to this space as the space of quadratically integrable functions on the
supergroup $G$ and will loosely denote it by ${\cal L}^2(G)$.
When we consider the space of functions on a supergroup, the group
invariant Laplacian, which is equal to the quadratic Casimir operator $C_2$,
acts on the space. The operator turns out to have a non-trivial Jordan form -- 
it is generically not diagonalizable \cite{Huffmann:1994ah,Schomerus:2005bf,Saleur:2006tf,Quella:2007hr}.

The first point we wish to make is that in our model the quadratic Casimir is
diagonalizable in cohomology.  When we impose the Siegel condition $b=0$ on
physical states, we will also need to impose that the quadratic Casimir $C_2$ annihilates
physical states. For simplicity, let's suppose first that the
quadratic Casimir $C_2$ has the form:
\begin{eqnarray}
C_2 &=& \left ( \begin{array}{cc} 0 & 1 \\
                                0 & 0 
              \end{array} \right)
\end{eqnarray}
in a certain sector of the state space. Its square is zero on this two-dimensional generalized
eigenspace.  It is then clear that the
state $ (0 \, \, \, 1)^{\mathrm{t}}$ will not be annihilated by $C_2$,
while on the space spanned by the state $ (1 \, \, \, 0)^{\mathrm{t}}$
alone, the quadratic Casimir {\em is} diagonalizable.  It is
diagonalizable in cohomology. This argument also holds when the
quadratic Casimir has a more elaborate Jordan form.

We think it is interesting to flesh
out this general observation
in a simple concrete example. This will give us the opportunity to
see how this simplification of the physical state space relates to other algebraic
properties of the state space. It will also allow us to find the classical origin
for the elimination of some generalized eigenvalue zero states from
the physical state space.

\section{A massless particle on the supergroup $GL(1|1)$} 
\label{gl11}
In this section, we will analyze in some detail the example of a particle
on the supergroup $G=GL(1|1)$. This simple example has the advantage that the
decomposition of the space of functions in terms of representations of
the left and right regular action of the group on itself is 
known \cite{Huffmann:1994ah,Schomerus:2005bf,Saleur:2006tf}, and not
difficult to rederive. All calculations can be done explicitly, and they
illustrate concretely the more advanced algebra that we will use in a later section.

The super Lie algebra
$\mathfrak{g}=gl(1|1)$ is an algebra that can be represented in terms of
 $2 \times 2$ supermatrices.  These supermatrices have bosonic
diagonal entries and fermionic  off-diagonal elements.
We can write the super Lie algebra in
terms of generators $h_{1,2}, e_1,f_1$ which satisfy the commutation
relations:
\begin{eqnarray}
\{ e_1,f_1 \} &=& h_2
\nonumber \\
{[} h_1,e_1] &=& 2 e_1
\nonumber \\  
{[} h_1,f_1] &=& -2 f_1,
\end{eqnarray}
while all other commutation relations are zero.  The fermionic
 annihilation and creation operators $e_1$ and $f_1$ anti-commute into
the central bosonic operator $h_2$. We also have an operator $h_1$
whose eigenvalue is raised or lowered when we 
annihilate or create a fermion

When we have in mind applications to backgrounds of string theory with
superconformal symmetries, it can be useful to think of the $gl(1|1)$
algebra as embedded into a superconformal algebra. 
 The generators of
 the $gl(1|1)$ subalgebra can then be identified with a subset of 
the superconformal generators and are more conventionally denoted as:
\begin{eqnarray}
h_2 &=&   \frac{1}{2} ( J-\Delta) = P_-
\nonumber \\
h_1 &=&  \Delta  +  J = 2 P_+
\nonumber \\
e_1 &=& Q^+
\nonumber \\
f_1 &=& Q^-,
\end{eqnarray}
where $\Delta$ measures the conformal dimension and $J$ the R-charge
while $ Q^{\pm} $ are supercharges of R-charge $\pm 1$ with correlated
 conformal 
dimension $ \pm 1$.
We note that the difference of the R-charge and the conformal
dimension is a central generator in this subalgebra. 
In many contexts, it is natural to take an exponentiation of the algebra
in which
the R-charge $J$ is the generator
of a compact $U(1)$ subgroup, while the scaling direction $\Delta$ is
taken to be non-compact.
A quadratic Casimir is given by the expression $C_2 =  \frac{1}{2} ( J^2-\Delta^2)
 + 2 Q^- Q^+$.
If we add any function of the central generator,
the resulting operator remains central.
\subsection{The classical foreshadowing}
\label{classical}
In this subsection, we will show that the reduction of the physical state space has
a counterpart in the classical theory.
For our concrete calculations, it is convenient to choose a matrix
realization of the group elements as follows (see e.g. \cite{Schomerus:2005bf}):
\begin{eqnarray}
g &=& e^{+i \sqrt{2} \eta_- Q^-} 
e^{i x^- P_- + i  x^+ P_+} e^{+i  \sqrt{2} \eta_+ Q^+ }
\nonumber \\
&=& \left( \begin{array}{cc} e^{\frac{i}{2} (-x^++ x^-)}+ \eta_+ \eta_- 
e^{\frac{i}{2}(x^+ +  x^-)} & 
i \eta_- e^{\frac{i}{2} ( x^+ +  x^-)} \\ i \eta_+ e^{\frac{i}{2} (x^+ +  x^-)} 
& e^{\frac{i}{2} ( x^+ +  x^-) }
 \end{array}\right),
\end{eqnarray} 
where $x^\pm$ are light-cone coordinates in $\mathbb{R}
\times \mathbb{R}$. {}\footnote{It is easy to adapt our analysis to the case where
a bosonic direction of the group
manifold is compact. } We have 
two Grassmann variables $\eta_\pm$.
The classical action for a  massless particle (in equation (\ref{action})) in this parameterization is:
\begin{eqnarray}
  S &=& \int  ( \partial_\tau x^+ \partial_\tau x^- 
+ 2 e^{ix^+}\partial_\tau \eta_+ \partial_\tau \eta_-) d \tau,
\end{eqnarray}
and it is supplemented with the constraint equation 
\begin{eqnarray}  \partial_\tau x^+ \partial_\tau x^- +
2 e^{ix^+} \partial_\tau \eta_+ \partial_\tau \eta_-
 &=& 0. \label{constraint}
\end{eqnarray}
When one solves the classical equations of motion, one finds that one needs
to distinguish two solution sets. The first solution set is parameterized by
integration constants $p_\pm, x^\pm_0 $ as well as $\pi_\pm, \eta_{\pm,0}$:
\begin{eqnarray}
x^+ &=& p_- \tau + x^+_0
\nonumber \\
\eta_\pm &=& \frac{i}{p_-} e^{-i p_- \tau-i x^+_0} \pi_{\pm}+ \eta_{\pm,0} 
\nonumber \\
x^- &=& -i \frac{2}{p_-^2} e^{-i p_- \tau-i x^+_0} \pi_{+} \pi_- + p_+ \tau+ x^-_0,
\end{eqnarray}
while the second one arises when the $x^+$-momentum $p_-$ is zero, and it
reads:
\begin{eqnarray}
x^+ &=& x^+_0
\nonumber \\
\eta_\pm
&=& e^{-i x^+_0} \pi_{\pm} \tau
+ \eta_{\pm,0} 
\nonumber \\
x^- &=&   2 i e^{-i x^+_0} \pi_+ \pi_- \frac{\tau^2}{2} + p_+ \tau + x^-_0.
\end{eqnarray}
The constraint equation (\ref{constraint}) on the first set reads:
\begin{eqnarray}
p_+ p_-  &=& 0,
\end{eqnarray}
or in other words the momentum $p_+=0$ (since $p_- \neq 0$ for the first set).
More interestingly, for the solution with zero light-cone momentum $p_-$, we find that
the constraint equation remains non-trivial:
\begin{eqnarray}
\pi_+ \pi_- &=& 0,
\end{eqnarray}
and we therefore find a second constraint, on top of the fact that
the momentum $p_-$ is zero.
Thus we find that when the momentum satisfies $p_-=0$, the space of classical solutions is smaller than when
$p_- \neq 0$.
We have two types of  trajectories. One is where the momentum $p_+$ is zero
and the momenta $p_-$ and $\pi_\pm$  (as well as $x^+_0,x^-_0,\eta_{\pm,0}$) are arbitrary.
The other type of trajectories is where the momentum $p_-$ is zero and the product of
$\pi_+$ with $\pi_-$ is also zero.

The origin of the surprising structure of the solution space is a
fermionic contribution to the length of the curve. We chose to
parameterize the curve by the proper time. The length of the curve is
always zero since we study a massless particle.  The length of the
curve is ordinarily the product of momenta $p_+ p_-$. However, when
the momentum $p_-$ is zero, we get a fermionic contribution to the
(generalized) length from the product $\pi_+ \pi_-$. We have to put
the latter combination to zero to obtain a curve of length zero.
In subsection \ref{BRST},
we will see that this reduction of the classical phase space is a
foreshadowing of the reduction of the state space in the quantum theory.

\subsection{The  space of functions on $GL(1|1)$}
In this subsection, we review properties of the 
space of quadratically integrable functions
on the supergroup $GL(1|1)$ \cite{Huffmann:1994ah,Schomerus:2005bf}.
 There is a
(delta-function normalizable) basis of the space of integrable functions
given by the exponentials:
\begin{eqnarray}
e_0(p_-,p_+) &=& e^{i p_- x^-+i p_+ x^+}
\nonumber \\
e_{\pm}(p_-,p_+) &=& e^{i p_- x^- + i p_+ x^+}  \eta_\pm
\nonumber\\
e_2(p_-,p_+) &=& e^{i p_- x^-+i p_+ x^+} \eta_- \eta_+.
\end{eqnarray}

\noindent
We denote the superspace functions with quadratically integrable
component functions by ${\cal L}^2(GL(1|1))$.  For future purposes, we review
the decomposition of the space of functions in terms of
representations of the left and right regular action of the group
\cite{Huffmann:1994ah,Schomerus:2005bf}.  The infinitesimal group actions on
the function space can be given in terms of differential operators.
The left-invariant vector fields act on the wave-functions as \cite{Schomerus:2005bf}:
\begin{eqnarray}
& & L_{P_-} = i \partial_{x^-} \qquad L_{P_+} = i \partial_{x^+}- \eta_+ \partial_+
\qquad
 L_{Q^+} = -i \partial_+ \qquad
L_{Q^-} = i e^{ix^+} \partial_- - \eta_+ \partial_{x^-},
\label{leftaction}
\end{eqnarray}
and the right invariant vector fields by:
\begin{eqnarray}
& & R_{P_-} = -i \partial_{x^-} \qquad R_{P_+} = -i \partial_{x^+} + \eta_- \partial_-
\qquad
R_{Q^+} = i e^{ix^+} \partial_+ + \eta_- \partial_{x^-} \qquad
R_{Q^-} = -i \partial_-. \label{rightaction}
\end{eqnarray}
In this example, the structure of the space  of functions can 
easily be determined by explicit calculation. It is useful to distinguish
the typical and the atypical sectors of the state space (where the nomenclature
originates in reference \cite{Kac:1977hp}). 
\subsubsection{Typical sector}
In the typical case where $p_- \neq 0$,  we have the
following group matrix elements corresponding to a direct summand $M_{\langle p_-,p_+ \rangle}$
in the decomposition of the quadratically integrable functions 
as a representation space under the left-right regular action \cite{Schomerus:2005bf}:
\begin{eqnarray}
M_{\langle p_-,p_+ \rangle}(g) &=& 
\left( \begin{array}{cc}
e^{ i p_- x^- + i (p_+-1) x^+} & i \eta_- e^{ i p_- x^- + i (p_+-1)x^+} \\
i p_- \eta_+ e^{ i p_- x^- + i (p_+-1)x^+} & p_- \eta_- \eta_+ e^{ i p_-  x^- + i (p_+-1) x^+}
+ e^{i p_- x^- +i p_+ x^+}
\end{array}
\right).
\end{eqnarray}
The functions in the first row of the above matrix form a basis of the
summand $H^R_{\langle p_-,p_+ \rangle}$ in the right regular
representation, where the space $H^R_{\langle p_-,p_+ \rangle}$ is a
typical graded representation space labeled by eigenvalues $\langle
p_-, p_+ \rangle$ (of one state -- the other state in the
representation has eigenvalues $p_-$ and $p_+-1$). This can be checked
by acting with the right generators given in equation
(\ref{rightaction}).  The second row forms a basis of the summand
$H^{R '}_{\langle p_-,p_+ \rangle}$ which is the same typical
representation, with opposite grading.
When we consider the left regular action, we note that it mixes the
two representation spaces with each other, to form a tensor product
representation $H^L_{\langle -p_-, -p_+ \rangle} \otimes H^R_{ \langle
  p_-, p_+ \rangle}$ of the left-right group actions. The part of the
function space with momentum $p_- \neq 0$ decomposes as a direct sum
of these tensor product representations. This type of structure is
familiar from the Peter-Weyl theorem for compact Lie groups.  If we
fix the momentum $p_-$ to be non-zero, we can draw a picture of the
action of the left and right generators on the summands of the state
space (see figure 1). In the diagram, we drop the common value of $-
i \partial_{x^-}$ from the notation, write the eigenvalue of
$-i \partial_{x^+}$ as a superscript and denote the function $p_-
e_2(p_-,p_+-1) + e_0(p_-,p_+)$ by $p_- e_2^{p_- -1}$. For ease of
illustration, we arbitrarily took the spacing between consecutive
values of $- i \partial_{x^+}$ to be one in the diagram.
\[
\xymatrixrowsep{2pc}
\xymatrix{
& &   &  & {e_+^{p_+-1}}  \ar@{-->}[ddd]<-1ex>_{R_{Q^+}} \ar[r]_{L_{Q^+}} & \ar@{-->}[ddd]<-1ex>_{R_{Q^+}} \ar@<-1ex>[l]_{L_{Q^-}} e_0^{p_+-1}   \\
& & {e_+^{p_+}}  \ar@{-->}[ddd]<-1ex>_{R_{Q^+}}  \ar[r]_{L_{Q^+}} &   {e_0^{p_+}} \ar@{-->}[ddd]<-1ex>_{R_{Q^+}}  \ar@<-1ex>[l]_{L_{Q^-}} & &
   \\ \ar@{-->}[ddd]<-1ex>_{R_{Q^+}}
 e_+^{p_++1}  \ar[r]_{L_{Q^+}}  & \ar@{-->}[ddd]<-1ex>_{R_{Q^+}} \ar@<-1ex>[l]_{L_{Q^-}} e_0^{p_++1}  & & && \\
&& & &  {p_- e_2^{p_+-1}}  \ar[r]_{L_{Q^+}}  \ar@{-->}[uuu]_{R_{Q^-}}  & 
\ar@<-1ex>[l]_{L_{Q^-}} {e_-^{p_+-1}} \ar@{-->}[uuu]_{R_{Q^-}}
   \\
& &  p_- e_2^{p_+ {\, \, }}   \ar[r]_{L_{Q^+}}  \ar@{-->}[uuu]_{R_{Q^-}} &    \ar@<-1ex>[l]_{L_{Q^-}} {e_-^{p_+}} \ar@{-->}[uuu]_{R_{Q^-}}  & &
   \\
 p_- e_2^{p_++1} \ar@{-->}[uuu]_{R_{Q^-}} \ar[r]_{L_{Q^+}}&  \ar@{-->}[uuu]_{R_{Q^-}} 
 \ar@<-1ex>[l]_{L_{Q^-}}e_-^{p_++1}  & & & &
}
\]
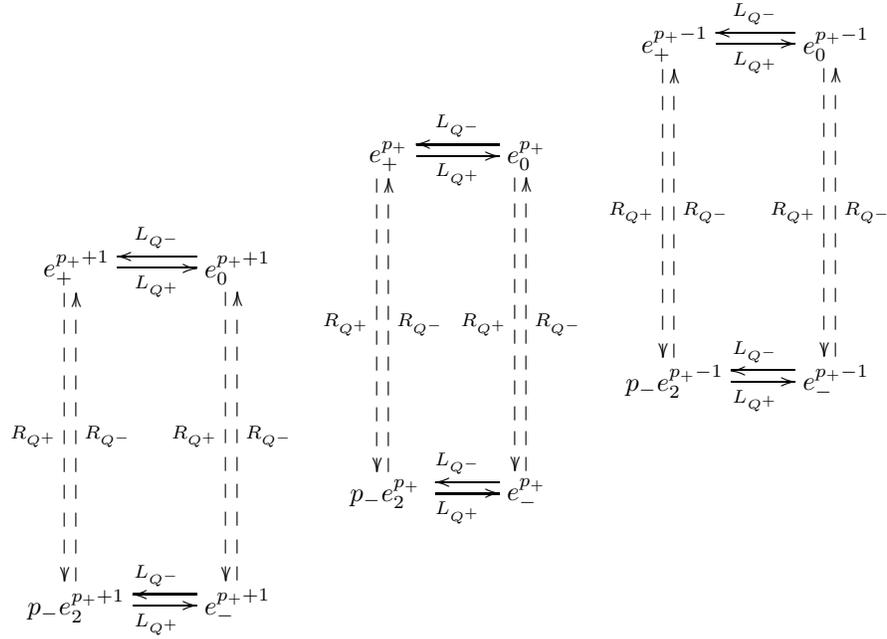
Figure 1:  The typical sector of the space of functions on $GL(1|1)$ as a representation of the left-right
regular action of the group. The left action is indicated with solid lines, the right action with stripes.

\vspace{.4cm}

The figure represents the fact that each summand contains
four states, that pair up two-by-two to form representations of either the left or the
right group action. The fermionic creation and annihilation operations are
invertible in this sector of the state space. The action of the quadratic
Casimir is diagonal and can be taken to be proportional to the product 
of lightcone momenta $p_- p_+$.


\subsubsection{Atypical sectors}
The structure of the state space is more interesting when the
lightcone momentum $p_-$ satisfies $p_-=0$. Because we concentrated on a
$\mathfrak{gl}(1|1)$ algebra, this condition has a particular
chirality.  It corresponds to demanding that the difference of twice
the conformal dimension and the R-charge is equal to zero.  In this
chiral subsector, the action of the generators on a basis of states is
as in the following diagram \cite{Huffmann:1994ah,Schomerus:2005bf}:
\[
\xymatrixrowsep{2pc}
\xymatrix{
& &   &  & {e_+^{p_+-1}}  \ar@{-->}[dl]_{R_{Q^+}} \ar[r]^{L_{Q^+}} &e_0^{p_+-1}   \\
& & {e_+^{p_+}}  \ar@{-->}[dl]^{R_{Q^+}} \ar[r]^{L_{Q^+}} & {e_0^{p_+}} & &
   \\
 e_+^{p_++1}  \ar[r]^{L_{Q^+}}  & e_0^{p_++1}  & & && \\
&& & &  {e_2^{p_+-1}} \ar@{-->}[dl]^{R_{Q^+}} \ar[r]^{L_{Q^+}}  \ar@{-->}'[u][uuu]_(.6){R_{Q^-}} \ar'[lu][lluu]_{L_{Q^-}} &{e_-^{p_+-1}} \ar@{-->}[uuu]_(.60){R_{Q^-}} \ar[lluu]_{L_{Q^-}} 
   \\
& & \ar'[lu][lluu]_(-.2){L_{Q^-}}  e_2^{p_+} \ar@{-->}[dl]^{R_{Q^+}} \ar[r]^{L_{Q^+}}  \ar@{-->}'[u][uuu]_(.6){R_{Q^-}} &    {e_-^{p_+}} \ar@{-->}[uuu]_(.40){R_{Q^-}} \ar[lluu]_{L_{Q^-}}  & &
   \\
 e_2^{p_++1} \ar@{-->}[uuu]_(.35){R_{Q^-}} \ar[r]^{L_{Q^+}}&  \ar@{-->}[uuu]_(.35){R_{Q^-}} e_-^{p_++1}  & & & &
}
\]
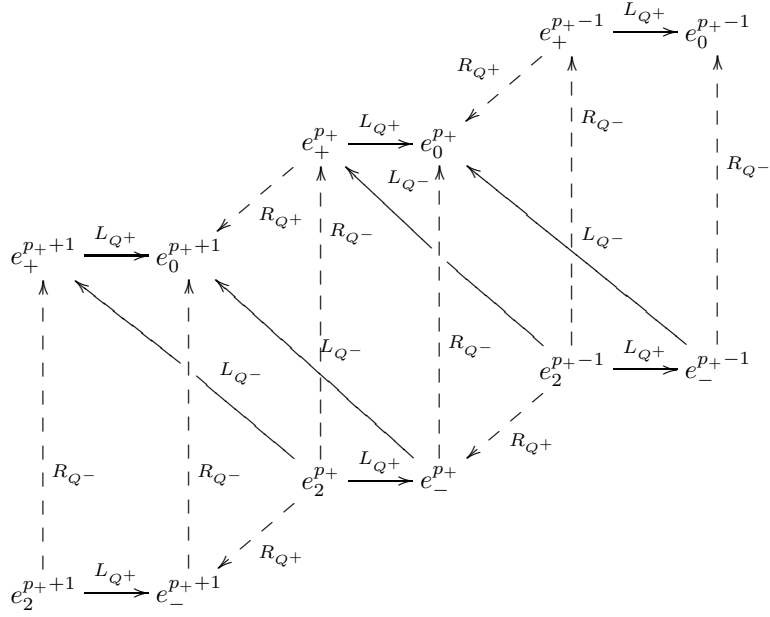
Figure 2: The left-right group action on functions on $GL(1|1)$ transforming in atypical representations.

\vspace{.4cm}

We see that the vectors $e_0^{p_+},e_+^{p_+-1},e_-^{p_+},e_2^{p_+-1}$ make up an indecomposable
right representation.
The vectors
$e_0^{p_+},e_+^{p_+},e_-^{p_+-1},e_2^{p_+-1}$ generate an indecomposable left 
representation. Note that the action of the fermionic
generators is no longer invertible.
The quadratic Casimir acts as 
the differential operator $2
e^{ix^+} \partial_- \partial_+$ in these representations, i.e. it maps
the state $e_2^{p_+}$  to the state $e_0^{p_++1}$ (for all momenta $p_+$) and it
annihilates all other states.
This sector forms an infinite dimensional indecomposable representation of the left-right
action of the group. We denote this summand of the representation space by $C_{p_+}$.


In summary, if we take the momentum $p_+$ to be continuous,
 the following decomposition for the space of quadratically integrable functions
holds \cite{Huffmann:1994ah,Schomerus:2005bf}:
\begin{eqnarray}
{\cal L}^2(GL(1|1))
&=& \int_{p_- \neq 0} \int dp_+ H^L_{\langle -p_- ,-p_+ \rangle}
\otimes H^R_{\langle p_-,p_+ \rangle} \oplus \int_0^1 dp_+ C_{p_+}.
\end{eqnarray}
The representation
space $C_{p_+}$ decomposes with respect to either the left or the right action
only as an infinite sum of projective four-dimensional representations
\cite{Huffmann:1994ah,Schomerus:2005bf}. Note that projective representations
 cannot be further extended without introducing
direct summands. The quadratic Casimir is diagonalizable, except in the atypical
sector. There it maps top states to bottom states, and is otherwise zero.

\subsection{The BRST cohomology}
\label{BRST}
In the quantum theory for the gauge fixed particle action, we tensor the
state space with a $(b,c)$ ghost
system. We take the latter to be a two-state system on which the quantum operators act as:
\begin{eqnarray}
b | \uparrow \rangle &=& | \downarrow \rangle, \qquad b | \downarrow \rangle = 0
\nonumber \\
c | \downarrow \rangle &=& | \uparrow \rangle, \qquad c | \uparrow \rangle = 0.
\end{eqnarray}

\subsubsection{Typical sector}
In the typical sector where $p_- \neq 0$, the quadratic Casimir acts diagonally with
eigenvalues  $p_+ p_-$. The cohomology localizes on momentum $p_+$
equal to zero. This is the familiar on-shell condition for a massless particle.

\subsubsection{Atypical sector}
When the momentum $p_-$ is zero,  the calculation of the cohomology
is more interesting.
In this case, the quadratic Casimir is zero on a large part of the
space, but it  maps top states $e_2^{p_+}$ to bottom states
$e_0^{p_++1}$.
The BRST closed states are the up states,
as well as the down states annihilated by the quadratic Casimir.
The BRST exact states are the up $e_0$ states.
Therefore, the closed non-exact states are the up states tensored with $e_\pm, e_2$,
and the $e_\pm, e_0$ down states.

The physical states satisfying the Siegel condition $b | \mbox{phys}
\rangle =0$, namely the down states, are the states
$e_{\pm},e_0$. Dually, the up states that are physical are
$e_\pm,e_2$. In many contexts, like the model of a particle in flat
space, where the Hamiltonian acts diagonally, the Siegel condition
lifts a two-fold degeneracy in the physical state space
cohomology. Here, because the quadratic Casimir is not diagonalizable,
this is not the case.  The cohomology on up states is dual to that on
down states (under a duality that maps all fermionic occupied states
to unoccupied states and vice versa).

\subsection*{The action of the quadratic Casimir}
In this example, we can see by inspection that the action of the
quadratic Casimir on the BRST
cohomology has become diagonal. This is a hands-on illustration of the general argument given in 
subsection \ref{diagonalizable}. We also saw  in 
subsection \ref{classical} that this feature has a classical counterpart.

\subsection{The action of space-time supersymmetry}
We now wish to point out a further algebraic property arising in this and more elaborate examples.
First of all, we note that the space-time isometries commute
with the BRST operator. The BRST cohomology is therefore again a representation
space of the supersymmetry algebra. What is the structure of this representation space?
\subsubsection{Typical}
In this sector, the cohomology coincides with the function space, and the representation
space of the left-right supercharges are ordinary tensor products of 
long multiplets (i.e. typical Kac modules).

\subsubsection{Atypical}
In the atypical sector, we will draw the representation space for the down states. The top
states are not present in the cohomology of down states. Removing them from the diagram of states gives
rise to the following supercharge actions on the physical state space:
\[
\xymatrixrowsep{1pc}
\xymatrixcolsep{3 pc}
\xymatrix{
& &   &  &
 {e_+^{n-1}}  \ar@{-->}[dl]_{R_{Q^+}} 
\ar[r]^{L_{Q^+}}  &e_0^{n-1}   \\
& & {e_+^{n}}  \ar@{-->}[dl]^{R_{Q^+}} \ar[r]^{L_{Q^+}} & {e_0^{n}} & &
   \\
 e_+^{n+1}  \ar[r]^{L_{Q^+}}  & e_0^{n+1}  & & && \\
&& & &  & {e_-^{n-1}} \ar@{-->}[uuu]_{R_{Q^-}} \ar[lluu]_{L_{Q^-}} 
   \\
& & &     {e_-^{n}} \ar@{-->}[uuu]_{R_{Q^-}} \ar[lluu]_{L_{Q^-}}  & &
   \\
 &  \ar@{-->}[uuu]_{R_{Q^-}} e_-^{n+1}  & & & &
}
\]
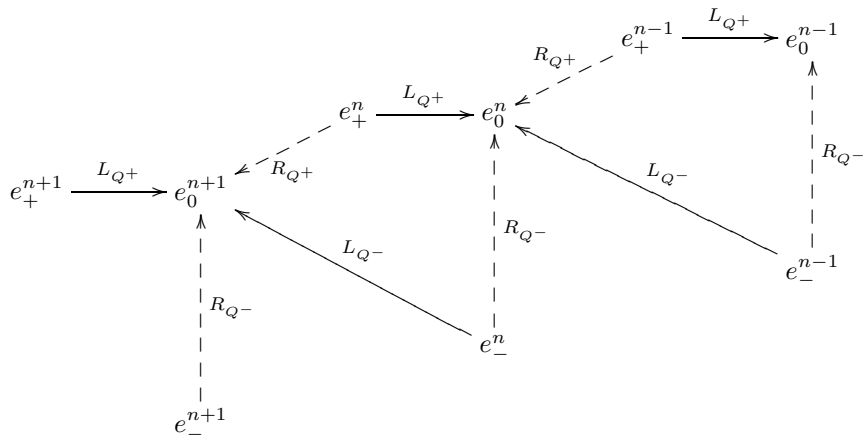
Figure 3: The left-right action of the superalgebra $\mathfrak{gl}(1|1)$ on the down physical
state space in the atypical sector.
\vspace{.4cm}

\noindent
We obtain a (non-unitary) indecomposable and infinite dimensional representation
space of the space-time super isometry algebra.
%

\subsection{Summary}
The non-diagonalizability of the Laplacian on the function space was
removed in the physical state cohomology. We were motivated to analyze this
phenomenon because in an embedding of the particle model in a conformal field theory,
the non-diagonalizability of the Laplacian on the function space
is inherited by the scaling
operator \cite{Schomerus:2005bf,Saleur:2006tf}, and it makes these
conformal field theories with supergroup targets logarithmic in nature. It is interesting to note
that worldline reparameterization invariance reduces the physical state space such
that the quadratic Casimir becomes diagonalizable.

In these particle models, we will still be left with a non-unitary
theory in cohomology.  We could define further space-time supercharge
cohomologies, on which the calculation of certain chiral correlation
functions localizes to unitarize these models.  Alternatively, one may
simply analyze further features of these interesting non-unitary theories.

In the following however, we will explore further how to reconcile these
models with expected properties of stringy space-time physics. The low-dimensional
example we studied up to now corresponds to the supersymmetrization of two light-cone
directions. To make progress, we consider an example with more space-time directions,
in which physical fluctuations  can survive. And we will study a more refined
cohomology that will alter the structure of the physical state space
more drastically than the cohomology associated to reparameterization invariance
alone.

\section{A massless particle on $AdS_3 \times S^3$}
\label{psu22}
A central building block in the Berkovits formulation of string theory
on $AdS_3 \times S^3$ with Neveu-Schwarz Neveu-Schwarz and
Ramond-Ramond fluxes is a conformal field theory with $PSU(1,1|2)$
supergroup target \cite{Berkovits:1999im}. The physical state space is
determined by computing a cohomology on a large space of conformal
field theory states \cite{Berkovits:1999im}\cite{Dolan:1999dc}. In
this section, we wish to solve for an important subset of the physical
state space of this model. It corresponds to the point particle limit
for string theory on $AdS_3 \times S^3$, in which we moreover
concentrate on compactification independent excitations. In other
words, we solve for the physical supergravity modes which correspond
to massless particle excitations on the supergroup. In this limit, the
subspace of physical states is determined by a cohomology on the space
of functions on the supergroup.  We will compute this cohomology, and
show how it simplifies the function space as a representation
space of the super isometry group. We also compare the physical states
we obtain to the result of lengthy calculations linking the Berkovits
formulation to supergravity \cite{Dolan:1999dc}, as well as the
Kaluza-Klein reduction of supergravity on $AdS_3 \times S^3$
\cite{Deger:1998nm}. The cohomological method for determining the
physical excitations, as well as their supermultiplet structure will
turn out to be efficient. In this section we will draw on more
advanced (super) algebra techniques which the reader can study for
instance from the references
\cite{Kac:1977hp}\cite{Zou}\cite{BrundanTilting}\cite{Gotz:2005ka}\cite{Drouot}.

\subsection{The space of functions on the group $PSU(2|2)$}
We will mostly work with the version of the supergroup target which
has a compact maximal bosonic subgroup. We consider the space of
quadratically integrable functions on $G=PSU(2|2)$. The super Lie
algebra $\mathfrak{g}$ corresponding to the group consists of four by
four hermitian matrices with zero trace and supertrace. To compute the
physical state space, it is useful to understand how the function
space decomposes into representations of the left and right regular
action of the group on itself.  We will start by analyzing the left
regular action.  To that end, we think of the function space as
consisting of a component function (which we take to be the top
component of a superfield on the supergroup) on the bosonic subgroup
$G_0=SU(2) \times SU(2)$ acted upon by all fermionic generators
through the left action of the group on itself. The function on the
bosonic subgroup can be decomposed by the Peter-Weyl theorem into
representations of $\mathfrak{g}_0=su(2) \oplus su(2)$ of the type
$\sum_{j_1,j_2=0,\frac{1}{2},\dots} M_{j_1,j_2}^L \otimes M^{\ast \,
  R}_{j_1,j_2}$ where $M_{j_1,j_2}^L$ is a representation of spins
$j_1$ and $j_2$ under the left action of the group on itself.
Therefore the space of functions on the supergroup splits into a sum
of representations of the left regular action of the type
\cite{Gotz:2006qp}:
\begin{eqnarray}
A = U(\mathfrak{g}) \otimes_{\mathfrak{g}_0} M^L_{j_1,j_2},
\label{A}
\end{eqnarray}
where $ U(\mathfrak{g})$ indicates the universal enveloping algebra of the super Lie algebra $\mathfrak{g}$.
We have kept the tensor product with the representation space of the right action of the bosonic
subgroup implicit for the moment.
By the Poincar\'e-Birkhoff-Witt theorem for the  universal enveloping
algebra $U(\mathfrak{g})$, 
the representation space $A$ consists of the states in the $su(2) \oplus su(2)$ representation,
acted upon by all eight fermionic operators.
The representation space $A$ has dimension $2^8 (2j_1+1) (2j_2+1)$. We would like to decompose it with
respect to the left action of the superalgebra.
\subsubsection{Representations of the algebra $psu(2|2)$}
In order to present the solution to this problem, we briefly review some Lie superalgebra representation theory 
(see e.g. \cite{Gotz:2006qp}).
We recall from the representation theory of $psl(2|2)$ that atypical Kac modules $K(j,j)=[j,j]$ (composed by
acting with all creation operators on a highest weight state with spins $(j,j)$) are composed
from short multiplets $L(j,j)=[j]$   through the diagram \cite{Gotz:2006qp}:
\[
\xymatrixrowsep{0pc}
\xymatrix{
& {[ } j+\frac{1}{2} {]}  \ar[rd] \\
[j] \ar[ru] \ar[rd] & & [j]   \\ 
& [ j - \frac{1}{2} ] \ar[ru]
}
\]
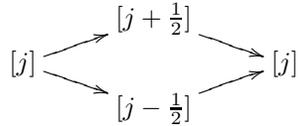
 Figure 4: The short multiplet composition series of atypical Kac modules
\vspace{.4cm}

\noindent
while a short multiplet $[j]$ contains the following $sl(2) \oplus sl(2)$ representations \cite{Gotz:2006qp}:
\begin{eqnarray}
[j]_{\mathfrak{g}_0} & \equiv & (j+1/2,j-1/2) \oplus 2 (j,j) \oplus (j-1/2,j+1/2).
\end{eqnarray}
If we would draw the action of fermionic generators on bosonic multiplets within a short representation,
it would be isomorphic to the diagram of the composition series of atypical Kac modules. We note that the diagrams
that we draw are strictly speaking only valid for $j$ larger than a particular lower bound at which exceptions
to the above diagrams occur. Those are not essential to our discussion, and we will ignore them throughout. 

The projective representations are the largest indecomposable covers
of these modules. To describe them, we need more mathematical results,
and we wish to discuss a subtlety in their description.  It has been
proven by algebraic means \cite{Zou,BrundanTilting} that for
representations of type I supergroups that satisfy that the
multiplicity of the simple quotient $L(\lambda)$ of a Kac module
$K(\lambda)$ in a composition series is no more than one, there is a
Bernstein-Gelfand-Gelfand reciprocity formula that holds. Namely, the
multiplicity of the Kac module $K(\lambda)$ in the projective
representation $P(\mu)$ associated to the highest weight $\mu$, is
equal to the multiplicity of the simple module $L(\mu)$ in the Kac
module $K(\lambda)$. We can state this briefly and roughly as the fact
that the Kac module is covered as it is composed.  In the case at
hand, there is a slight complication which is (as we saw) that the
multiplicity of the short representation in the composition series of
the atypical Kac module is not equal to one, and that therefore the
theorem quoted above cannot be applied directly. We 
circumvent this complication by lifting the $psl(2|2)$ representation
to a representation of the algebra $gl(2|2)$. In other words, we
provide the representation space with an extra grading that keeps
track of the number of fermionic creation minus the number of
fermionic annihilation operators that act on a ground state.  Thus we
lift the degeneracy of the short multiplet in the composition series,
and we can apply the result on the multiplicities of Kac modules in
the projective cover to confirm that the projective representation
$P(j,j)$ is composed out of Kac modules as described in
\cite{Gotz:2006qp}:
\[
\xymatrixrowsep{0pc}
\xymatrix{
& {[ } j+\frac{1}{2},j+\frac{1}{2} {]}  \ar[rd] \\
[j,j] \ar[ru] \ar[rd] & & [j,j]   \\ 
& [ j - \frac{1}{2},j-\frac{1}{2} ] \ar[ru]
}
\]
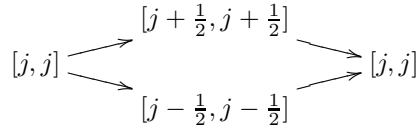
Figure 5: The Kac composition of the projective representation $P(j,j)$.
\vspace{.4cm}

\noindent
Moreover, we can use the results on how projective representations of the algebra $gl(2|2)$ are
composed of short multiplets \cite{Drouot} to rederive that the projective representation of $P(j,j)$ has the
following structure \cite{Gotz:2006qp}:
\[
\xymatrixrowsep{0pc}
\xymatrix{
                                    & & [j+1] \ar[rd] & & \\
& 2 {[ } j+\frac{1}{2} {]}  \ar[rd] \ar[ru] & & 2 {[} j + \frac{1}{2} ] \ar[rd] & \\
[j] \ar[ru] \ar[rd] & & 4 [j] \ar[ru] \ar[rd]  && [j] \\ 
& 2 [ j - \frac{1}{2} ] \ar[ru] \ar[rd] & & 2 [ j - \frac{1}{2} ] \ar[ru] & \\
                                     & & [j-1] \ar[ru] & & 
}
\]
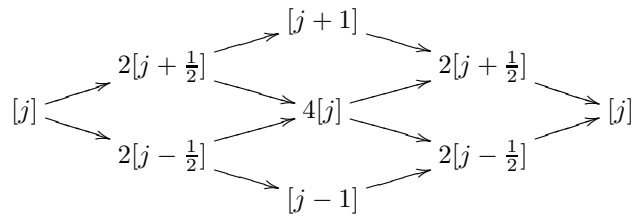
Figure 6: The composition of the projective representation $P(j,j)$ in terms of short multiplets.
\vspace{.4cm}

\noindent
For our purposes, if will be useful to have more explicit information about the grading. It can
be gleaned from the results of \cite{Drouot} which we recall in appendix \ref{gl22}
that the additional $u(1)$ grading that an embedding in $gl(2|2)$ provides will partially
lift the degeneracies in the above diagram to:

\[
\xymatrixrowsep{0pc}
\xymatrix{
                                    & & [j+1]_0 \ar[rd] & & \\
&  {[ } j+\frac{1}{2} {]}_{+1} \oplus  {[ } j+\frac{1}{2} {]}_{-1}  \ar[rd] \ar[ru] & &  
{[ } j+\frac{1}{2} {]}_{-1} \oplus  {[ } j+\frac{1}{2} {]}_{+1} \ar[rd] & \\
[j]_0 \ar[ru] \ar[rd] & & 2 [j]_0 \oplus [j]_{-2} \oplus[j]_{+2} \ar[ru] \ar[rd]  && [j]_0 \\ 
&  [ j - \frac{1}{2} ]_{+1} \oplus [j-\frac{1}{2} ]_{-1} \ar[ru] \ar[rd] & & [ j - \frac{1}{2} ]_{-1} \oplus [j-\frac{1}{2} ]_{+1} \ar[ru] & \\
                                     & & [j-1]_0 \ar[ru] & & 
}
\]
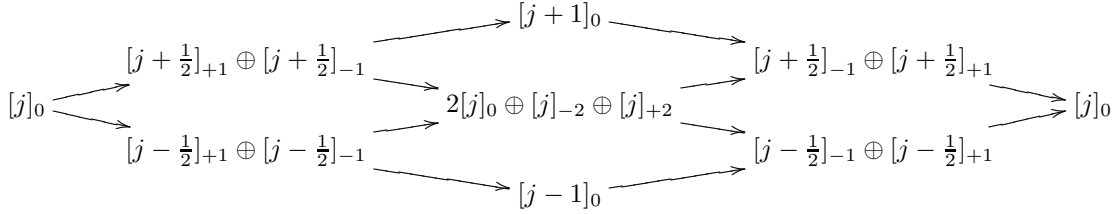
Figure 7: The graded composition of the projective representation $P(j,j)$ in terms of short multiplets.

\vspace{.4cm}
\noindent
This result is also coded in the action of the
outer automorphisms of $psl(2|2)$ on the short multiplets \cite{Gotz:2005ka}.
An intuitive reading of the diagram goes as follows. In the projective
representation $P(j,j)$, we have two
annihilation operators that act non-trivially on a generating vector
 on the left, to generate two new top representations in
Kac composition factors (of charge $-1$ say). When both act consecutively, we have the
fourth and last top representation in a Kac composition factor (of charge $-2$). All
other short representations are obtained through the action of
fermionic creation operators (of charge $+1$) within a given Kac composition factor.

The method we used in this subsection to lift $sl(n|n)$ or $psl(n|n)$ representations in order to 
be able to apply results for $gl(n|n)$ works generically.

\subsubsection{The left regular representation on the superfield}
With these prerequisites in hand, we can argue how the representation space $A$ of
equation (\ref{A}) contained within a superfield $V$
on the supergroup decomposes with respect to the left regular action. First of all, it is known \cite{Zou}
that the representation space $A$ permits a Kac composition series. Moreover, it can be reconstructed
as in the proof of Lemma 2.3 of \cite{Zou}. In short, the Kac composition factors
correspond one to one to the representations of the bosonic subalgebra appearing
in a Kac module.
When the two spins $j_{1,2}$ of the bosonic representation space
$M^L_{j_1,j_2}$ on which the representation $A$ is built are equal,
$j_1=j_2$, we have eight Kac composition modules that are atypical
which combine four by four into two projective representations $P(j_1,j_1)$. When
the spins satisfy $j_1=j_2+1$, we have four Kac composition modules that are atypical
which combine into one projective representation $P(j_1-1/2,j_1-1/2)$, and for
spins $j_1=j_2-1$ we obtain the projective representation
$P(j_1+1/2,j_1+1/2)$. All other Kac composition modules that appear in the representation
space $A$ are typical (and therefore projective). They are direct summands in the representation
$A$. That characterizes fully the representation space $A$.

When we tensor back in the right representation space $M^{\ast \,
  R}_{j_1,j_2}$, and concentrate on the atypical summands in the
superfield $V$, we obtain the atypical part of the space of functions:
\begin{eqnarray}
V_{atyp} & = & 
\sum_{j_1=0}^{\infty} P(j_1,j_1)_L \otimes (2 (j_1,j_1)_R \oplus (j_1+1/2,j_1-1/2)_R \oplus (j_1-1/2,j_1+1/2)_R).
\end{eqnarray}
The right representations of the Lie algebra 
$\mathfrak{g}_0$ necessarily combine into a right short multiplet:
\begin{eqnarray}
V_{atyp} & = & 
\sum_{j_1=0}^{\infty} P(j_1,j_1)_L \otimes  [j_1]_R.
\label{Ptimesshort}
\end{eqnarray}
The formula gives the decomposition of $V_{atyp}$ under the left
regular action (and not under the full right regular action as we will see
in detail).
Again, we remind of the caveat that these results are strictly valid
only for spins $j_1$ large enough. We only wrote the atypical part of
the representation space since we will later be interested in the
space of states with quadratic Casimir generalized eigenvalue equal to
zero. The typical part decomposes as in the Peter-Weyl theorem for compact groups.

\subsubsection{The big picture}
We have fully characterized the left regular representation on the
space of functions.  We want to combine it into one big picture with
the action of the right generators on the representation space. The
picture is big since we have sixteen short multiplets in each
indecomposable representation on the left, and because the right
action further mixes left representation spaces. A partial diagram of
the left and right actions on left-right short multiplets that compose
the representation space in the atypical sector is sketched in figures
8 and 9. We drew the left projective representations (with full
lines), and (a small part of) the right action in striped
lines. Figure 8 is a detail of figure 9. In the second figure, we left
out the labeling of the grid by tensor products of left-right short
multiplets. {From} these two pictures one can reconstruct the diagram
extending towards higher and lower spins. The picture degenerates near
spin zero (in a way that can be derived from the results in
\cite{Drouot}). It is straightforward to further split and grade the
picture with an additional $u(1)$ left and $u(1)$ right grading (as we
did in figure 7). We invite the reader to picture the resulting
diagram.


\xymatrixrowsep{2.5pc}
\xymatrixcolsep{2pc}
\xymatrix @u {
           &&  [ j - \frac{1}{2} ] \otimes  [ j - \frac{1}{2} ] \ar[rd] \ar[ld] \ar@{-->}[rrrrrrd] &&& 
                        & & [j] \otimes [j] \ar[rd] \ar[ld] & & \\
           & 2 {[ } j {]} \otimes [ j - \frac{1}{2} ] \ar[rd] \ar[ld] &&  
2 [ j - 1 ] \otimes [j - \frac{1}{2} ] \ar[rd] \ar[ld] & &
&  2 {[ } j+\frac{1}{2} {]} \otimes [j] \ar[rd] \ar[ld] & &  
   2 [ j - \frac{1}{2} ]  \otimes [j] \ar[rd] \ar[ld] \ar@{-->}[lllllld] & \\
       [ j + \frac{1}{2} ] \otimes  [ j - \frac{1}{2} ] \ar[rd]      && 
4 [ j - \frac{1}{2} ] \otimes  [ j - \frac{1}{2} ] \ar[ld] \ar[rd] \ar@{-->}[rrrrrrd] &&  
  [ j - \frac{3}{2} ] \otimes  [ j - \frac{1}{2} ] \ar[ld]  &
    [j+1]\otimes [j]  \ar[rd] & & 4 [j]\otimes [j]  \ar[rd] \ar[ld] && [j-1]\otimes [j] \ar[ld] \\ 
           &2 {[ } j {]} \otimes [ j - \frac{1}{2} ] \ar[rd] && 2 [ j - 1 ] \otimes [j - \frac{1}{2} ]  \ar[ld]&&
& 2  [ j + \frac{1}{2} ] \otimes [j]  \ar[rd] & &
 2 [ j - \frac{1}{2} ]\otimes [j] \ar[ld] \ar@{-->}[lllllld]  & \\
           &&  [ j - \frac{1}{2} ] \otimes  [ j - \frac{1}{2} ] &&&
                                     & & [j] \otimes [j] & & 
}

\noindent
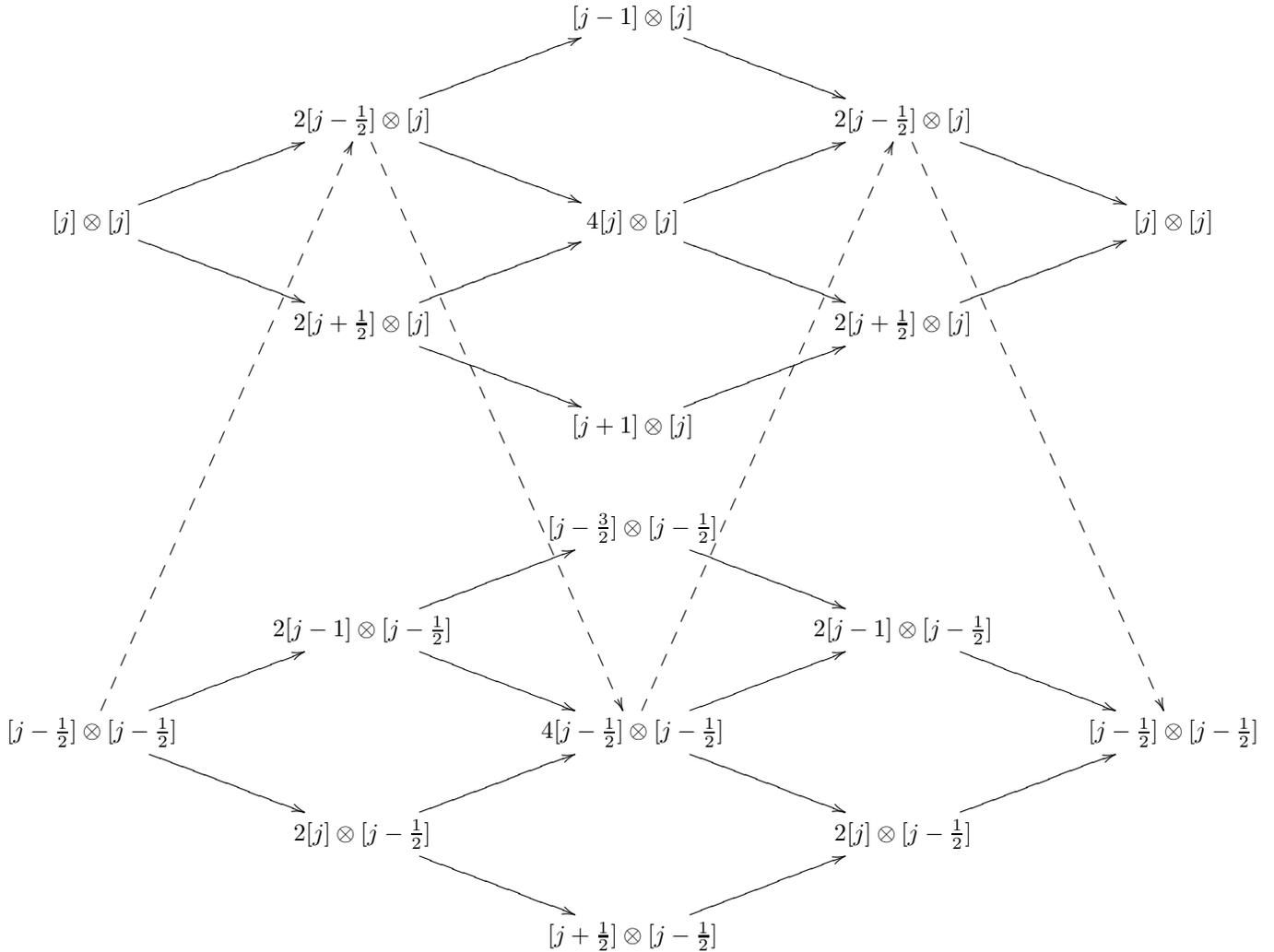
Figure  8: Detail of the decomposition of the atypical sector of the function space as a sum of left projective
representations interconnected by the right group action.
\vspace{.4cm}




\begin{center}
\xymatrixrowsep{2pc}
\xymatrixcolsep{1pc}
\xymatrix  {
           &&  \ar[rd] \ar[ld]  &&     & 
           &&  \ar[rd] \ar[ld] && &  
           &&\ar[rd] \ar[ld] \ar@{-->}[rrrrrrd]\ar@{-->}[lllllld]&& & 
           && \ar[rd] \ar[ld] && & 
           && \ar[rd] \ar[ld] &&\\
           & \ar[rd] \ar[ld] &&   \ar[rd] \ar[ld] &&
&   \ar[rd] \ar[ld]\ar@{-->}[lllllld] \ar@{-->}[rrrrrrd] &&      \ar[rd] \ar[ld]  &&
& \ar[rd] \ar[ld]&&\ar[rd] \ar[ld]&&
&\ar[rd] \ar[ld]&& \ar[rd] \ar[ld]\ar@{-->}[lllllld] \ar@{-->}[rrrrrrd] &&
& \ar[rd] \ar[ld]&&\ar[rd] \ar[ld]&\\
      \ar[rd]    \ar@{-->}[rrrrrrd]  && \ar[ld] \ar[rd] & &    \ar[ld]  &
    \ar[rd] & & \ar[rd] \ar[ld] && \ar[ld]  & 
\ar[rd]&&\ar[rd] \ar[ld]\ar@{-->}[lllllld] \ar@{-->}[rrrrrrd]&&\ar[ld] & 
\ar[rd]&&\ar[rd] \ar[ld]&& \ar[ld] & 
\ar[rd]&&\ar[rd] \ar[ld]&&\ar[ld]\ar@{-->}[lllllld] \\ 
           & \ar[rd] &&   \ar[ld]&&
&   \ar[rd]\ar@{-->}[rrrrrrd] & & \ar[ld]   &  &
&\ar[rd] &&\ar[ld]&&
&\ar[rd]&&\ar[ld]\ar@{-->}[lllllld] &&
&   \ar[rd] & & \ar[ld]   &  \\
           &&&&&
                                     & &  && &  && && && && & &&&&& 
}

\end{center}
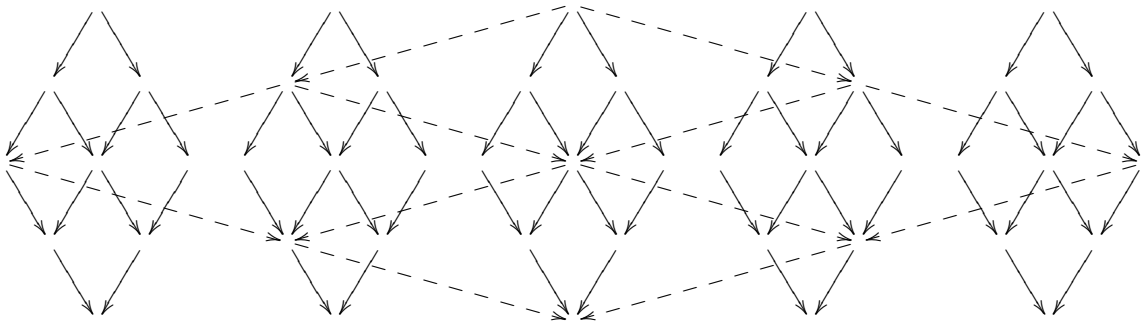
Figure 9: A selection and a slice out of an even bigger picture, showing how one
(striped) right projective representation interconnects with (solid)
left projective representations in the atypical sector of the space of
functions on the supergroup. The figure has been rotated ninety degrees clockwise
with respect to the previous one.

\subsection{The cohomology defined}
We have understood the structure of the space of functions on the
supergroup, and can now determine which states are physical.  The
space of physical particle states which are compactification
independent is obtained by imposing constraints. These constraints
were derived in the Berkovits formulation of string theory on $AdS_3
\times S^3$ with Neveu-Schwarz Neveu-Schwarz and Ramond-Ramond fluxes
\cite{Berkovits:1999im}.  First of all, it was argued that the
physical cohomology is coded in a single function $V$ on the
supergroup \cite{Berkovits:1999im}. The square of the quadratic
Casimir should vanish on the function\footnote{That does not imply
  that the quadratic Casimir is zero, since it is not
  diagonalizable.}. Thus we can restrict our analysis to generalized
eigenspaces of eigenvalue zero.

It is convenient to express the further constraints in terms of
generators that make the $so(4)$ representation content of the adjoint
of the $psl(2|2)$ algebra manifest. We can take generators such that
they satisfy the commutation relations \cite{Berkovits:1999im}:
\begin{eqnarray}
{[} K_{ab}, K_{cd} {]} &=&  \delta_{ac} K_{bd} + \delta_{bd} K_{ac} - \delta_{ad} K_{bc} - \delta_{bc} K_{ad}
\nonumber \\
{[} K_{ab} , F_c {]} = \delta_{ac} F_b - \delta_{bc} F_a & , & {[} K_{ab} , E_c {]} = \delta_{ac} E_b - \delta_{bc} E_a
\nonumber \\
\{ E_a , F_b \} &=&  \frac{1}{2} \epsilon_{abcd} K^{cd},
\label{psl22}
\end{eqnarray}
while all other commutators are zero.  The index $a \in \{ 1,2,3,4 \}$
is an $so(4)$ vector index and the bosonic generators $K_{ab}$ are in
the (anti-symmetric) adjoint.  The further constraint equations on the
superfield $V$ derived in \cite{Dolan:1999dc} are that $F^4 V=0$ as
well as $K_{ab} F^a F^b V=0$. Moreover, the functions of the form
$K_{ab} F^a F^b W $ are gauge trivial.  These constraints are valid
for both the left and the right actions on the function space, and are
moreover $psl(2|2)$ covariant in a subtle way spelled out in
\cite{Dolan:1999dc}.  Below we will concentrate on the cohomology of
the operator $K_{ab} F^a F^b$ in the space of generalized
eigenfunctions of eigenvalue zero.  All other left constraints will
then automatically be satisfied in this cohomology.

We will see that on the cohomology of the operator $K_{ab} F^a F^b$
the quadratic Casimir vanishes. In particular, this implies that the
model will be reparameterization invariant (as for the massless
particle on the supergroup in section 2). However, the string cohomology is
more refined and in particular it will also eliminate some unphysical
fermionic directions in space-time. The underlying idea is that the string
cohomology must arise from a model which also has fermionic reparameterization
invariances.

\subsection{The left cohomology in a projective summand}
First we analyze the cohomology for the constraints associated to the
left action of the group on itself.  Since the generalized eigenspaces
of eigenvalue zero correspond to a sum of atypical projective modules for the
left action, we will work in one direct summand projective module.
The constraint for a state in the projective module to be physical is
$K_{ab} F^a F^b | \mbox{phys} \rangle =0$. All states in the
projective module can be generated from a single state. It is a
highest weight state of spins $(j,j)$ with respect to the bosonic
subalgebra $\mathfrak{g}_0=sl(2) \oplus sl(2)$, and, as we saw before, it can be acted upon
by up to two fermionic annihilation operators $E$ to give new top states for Kac
composition factors. On those states we can act with any number of fermionic creation
operators $F$ to fill out a Kac module. We analyze the physical state condition
level by level in the number of fermionic creation operators $F$ acting on the generator of
the module. Here, a top state is a state at level $F^0$.

We have that states obtained by the action of four creation operators
$F$ satisfy the constraint automatically, as do states at level
$F^3$. When two creation operators $F$ act, we must take into account
the following facts.  The constraint equation $K_{ab} F^a F^b |
\mbox{phys} \rangle =0$ is scalar in terms of the bosonic
subalgebra. It generates $(2j'+1)^2$ independent constraints in Kac
modules built on $(j',j')$ representations. In other words, for each
Kac composition factor in the projective module, the constraint
equations eliminate one (bosonic) $(j',j')$ representation at the
middle level. We are left with states in the $(2-1)(j',j') \oplus
(j'\pm 1,j') \oplus (j', j' \pm 1)$ $\mathfrak{g}_0$-representations
that satisfy the constraint equation at level $F^2$, in each composite
Kac module of spin $(j',j')$. For future purposes, we note that the
states $K_{ab} F^a F^b$ acting on the top state in any Kac composition
factor satisfy the constraint automatically.  That is because the
constraint acting on such a state gives rise to the bosonic quadratic
Casimir operator.  The bosonic quadratic Casimir evaluated on a top
state in an atypical Kac module is zero.

The analysis at first order in the operators $F$ is a little more
intricate. As an intermediate step, it will be useful to compute the action of
the bosonic quadratic Casimir on a generic state at level $F^3$. Since the
bosonic quadratic Casimir $C_2^{bos}$ satisfies the following relation
with the total quadratic Casimir $C_2^{tot}$: $ E_a F^a = C_2^{tot}
- C_2^{bos}$, we will start by computing the action of $E_a F^a$ on
a state at level three:
\begin{eqnarray}
E_e  F^e  c_d \epsilon^{abcd} F_a F_b F_c | \mbox{top} \rangle  &=& 
\nonumber \\
(3 c^{[ a} K^{bc ]} F_a F_b F_c 
+   \frac{1}{4}  c^d \epsilon^{abce} F_e F_b F_c F_a E_d)  | \mbox{top} \rangle &=& 
\nonumber \\
(3 c^{[ a} K^{bc ]} F_a F_b F_c 
- c_a \epsilon^{abce} F_e F_b F_c E^d F_d)  | \mbox{top} \rangle &=& 
\nonumber \\
(3c^{[ a} K^{bc ]} F_a F_b F_c 
+  c_a \epsilon^{abce} F_e F_b F_c  (C_2^{bos} - C_2^{tot}) ) | \mbox{top} \rangle &=& 
\nonumber \\
(3 c^{[ a} K^{bc ]} F_a F_b F_c 
+ C_2^{tot}  c_d \epsilon^{abcd} F_a F_b F_c) | \mbox{top} \rangle. &&
\end{eqnarray}
We conclude that we have that the operator $C_{2}^{bos} = C_2^{tot} - E^a
F_a$ acting on a state at level three is zero if and only if $c^a
K^{bc} F_a F_b F_c | \mbox{top} \rangle$ is equal to zero. This
implies that there is a state $c^a F_a | \mbox{top} \rangle$
at level one which satifies the constraint equation for every state at
level three whose bosonic quadratic Casimir is zero. That implies that
the physical states at level one  in a Kac composition factor built on top
states with spin $(j',j')$ are the $sl(2) \oplus sl(2)$ representations $(j'+1/2,j'+1/2)$
and $(j'-1/2,j'-1/2)$. Finally, at level zero, there are no solutions
to the constraint equation. Thus, we have found all closed states.

The gauge trivial or $K^{ab} F_a F_b$ exact states are found as
follows. At level $F^4$, all states are gauge trivial. At level $F^3$,
we use again the calculation above that says that we can reach all
level three states whose bosonic quadratic Casimir is non-zero. We are
left with the states $(j' \pm 1/2, j' \pm 1/2)$ at level three. At
level two, the states $K_{ab} F^a F^b | \mbox{top} \rangle$ are gauge
trivial, and form a $(j',j')$ representation which is different from
the one excluded by the physical constraint condition (as follows by
the remark made previously on the states at level two). Thus, in each
Kac composition factor $[j',j']$, we are left with the representation content
of the two middle short multiplets
$[j' \pm 1/2]$. We apply this reasoning to all Kac composition factors in a
left projective module and find that starting from figure 7,
we are left with a representation
of the left $psl(2|2)$ action on the cohomology described by figure 10:
\begin{center}
\[
\xymatrixrowsep{0pc}
\xymatrix{
                                    & & [j+1]_0 \ar[rd] & & \\
&  {[ } j+\frac{1}{2} {]}_{+1}  \ar[rd] \ar[ru] & &  
 {[ } j+\frac{1}{2} {]}_{-1}  & \\
 & & 2 [j]_0  \ar[ru] \ar[rd]  &&  \\ 
&  [j-\frac{1}{2} ]_{+1} \ar[ru] \ar[rd] & & [j-\frac{1}{2} ]_{-1} & \\
                                     & & [j-1]_0 \ar[ru]  & & 
}
\]
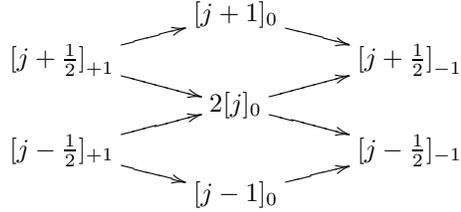
{Figure 10: The left cohomology in a graded atypical projective module.}
\end{center}
\subsection{The full cohomology}
We have just computed the cohomology with respect to the generators of
the left action of the supergroup on itself. We now need to further
compute the cohomology with respect to the right action of the
supergroup. Since the cohomological operators commute (since the left
and the right action of the supergroup on itself commute), we can
compute the cohomologies independently, and then restrict to the
representations which are non-trivial in both complexes.

Since the right action of the group on itself is isomorphic to the left
action of the group on itself,
we have a very similar answer for the right-moving cohomology in the right 
atypical projective modules.
There is one important difference, which is that we assign the
opposite grading to the right fermionic creation and annihilation
operators\footnote{This is dictated for instance by the demand that
  one recuperates flat space supergravity in the infinite radius
  limit.}.  To make this concrete, let's first define the algebra of
generators of the right action of the group on itself to be again
a $psl(2|2)$ algebra as in equations (\ref{psl22}). We denote
all of them with an extra bar.
 
The right cohomology is now taken with respect to an
operator $\bar{K}_{ab} \bar{F}_a \bar{F}_b$ of opposite $u(1)$ grading.
Thus, where the representations $[j\pm 1/2]_{+ 1}$ survived in the
left cohomology, the representations $[j \pm 1/2]_{-1}$ will survive
in the right cohomology, and vice versa. The resulting right cohomology
in a graded atypical right projective module will be:
\begin{center}
\[
\xymatrixrowsep{0pc}
\xymatrix{
                                    & & [j+1]_0 \ar[rd] & & \\
&  {[ } j+\frac{1}{2} {]}_{-1}  \ar[rd] \ar[ru] & &  
 {[ } j+\frac{1}{2} {]}_{+1}  & \\
 & & 2 [j]_0  \ar[ru] \ar[rd]  &&  \\ 
&  [j-\frac{1}{2} ]_{-1} \ar[ru] \ar[rd] & & [j-\frac{1}{2} ]_{+1} & \\
                                     & & [j-1]_0 \ar[ru]  & & 
}
\]
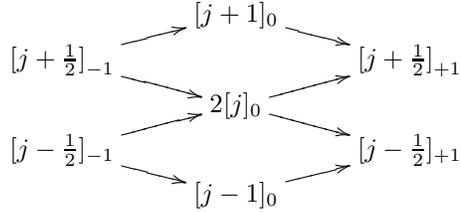
{Figure 11: The right cohomology in a graded atypical projective module.}
\end{center}

Therefore, in each projective module, after taking both left and right
cohomologies into account (combining figures 10 and 11 with figures 8
and 9), we will only be left with the middle short multiplets $[j+1]_0
\oplus 2 [j]_0 \oplus [j-1]_0$ of zero grading. Indeed, the grade $+1$
representations one removed from the top level are eliminated by the
right cohomology (see figure 11) while the grade $-1$ representations
one removed from the top level are eliminated by the left
cohomology (see figure 10). A similar reasoning, exchanging left and right
cohomologies in the argument, shows that the level one removed from
the bottom level is also entirely eliminated in the double complex.

The full solution to our
cohomological problem is then a sum over the spin $j$ of the
representations $([j+1]_L \oplus 2 [j]_L \oplus [j-1]_L ) \otimes
[j]_R$, where we tensored in the right short multiplet of equation
(\ref{Ptimesshort}).  In conclusion, we found the physical state
space:
\begin{equation}
{V}_{ phys} = 
\sum_{j=0}^\infty ([j+1]_L \otimes [j]_R \oplus 2 [j]_L \otimes [j]_R
\oplus [j]_L \otimes [j+1]_R),
\end{equation}
where we have written the solution in a manifestly left-right
symmetric manner.  We note that the full cohomology has reduced to a
direct sum of tensor product spaces of short representations of the
left and right supersymmetry algebra. The big infinite dimensional
indecomposable structure has been cut into finite dimensional and
unitary representations by the sharp scissors of physical cohomology.

\subsection*{Remarks}
Since the final result localized on middle short multiplets in the Kac
composition factors, it should be clear that the quadratic Casimir
itself vanishes in cohomology (since the quadratic Casimir acts
diagonally up to a term that changes the level of short multiplets).
All constraints on physical states are automatically satisfied once we
restrict to the above left-right cohomology.

An important difference with the reparameterization invariant superparticle
on $GL(1|1)$ is the fact that the physical cohomology consists of finite
dimensional representations of the supergroup. The origin of this further
reduction lies in the fact that the Berkovits cohomology is more refined,
and in particular eliminates all fermionic target space directions, rendering
the model unitary. To obtain a similar finding in the $GL(1|1)$ case,
one would need to refine the cohomology beyond the quadratic Casimir
operator, for instance by introducing
 a BRST operator proportional to a space-time supercharge.

\subsection{The comparison with Kaluza-Klein supergravity results}
We can compare our final answer to two related results in the
literature.  Firstly, in \cite{Dolan:1999dc} it was shown that the
physical state conditions agree with the linearized supergravity
equations of motion, by explicitly realizing the action of the
symmetry algebra as differential operators acting on the component
fields. Secondly, in \cite{Deger:1998nm} the Kaluza-Klein reduction of
$(2,0)$ chiral supergravity on $AdS_3 \times S^3$ 
 was performed in terms of the component fields. 
The final result of this two-step analysis of physical states
can be seen in  figures 1, 2 and 3 in \cite{Deger:1998nm}.
In our compact notation, the figures 2 and 3 correspond to two
 $[j]_L \times [j]_R$
representations of the algebra $psu(2|2)$ {}\footnote{We have that $n=1$ in \cite{Deger:1998nm} since we only
  have a single tensor multiplet in our supergravity theory
  \cite{Berkovits:1999im}.}.
Similarly, by rendering the $su(2) \oplus su(2)$
representation content of the multiplets $[j-1]_L \times [j]_R$ and
$[j]_L \times [j-1]_R$ manifest, we can match them onto the multiplet
visualized in figure 1 of \cite{Deger:1998nm}, and its conjugate
multiplet. We have found full agreement.

In passing we note that the technique used in
\cite{Deger:1998nm} of comparing Kaluza-Klein reduction on a sphere to
Kaluza-Klein reduction on $AdS_3$, by analytic continuation, precisely agrees
with the analytic continuation technique used here. We claim therefore
that the analysis in the case of $PSU(1,1|2)$
runs along precisely the same lines as the analysis performed in this
paper. The crucial technical aspect of the analysis will be that the weight
spaces of the representations that arise are all finite dimensional.
It will be interesting to confirm this expectation by explicitly 
analyzing the extension of the results of \cite{Drouot} on the structure
of projective representations to the case of projective representations
built on discrete lowest and highest weight representations, and to 
carefully state the mathematical and cohomological 
results in the context of the category
of representations with finite dimensional
weight spaces.

In conclusion, we observe that we not only coded the supergravity
equations of motion \cite{Dolan:1999dc} algebraically, but we also
immediately obtained their solution upon Kaluza-Klein reduction of
supergravity on $AdS_3 \times S^3$ \cite{Deger:1998nm}. By keeping the
space-time super isometries manifest, we were able to calculate very
efficiently.  We thus reaped a reward for working in the Berkovits
formalism.

\section{Conclusions}
The space of functions on a supergroup has a quadratic Casimir, or
Laplacian, with non-trivial Jordan form. That property is inherited by
conformal two-dimensional sigma-models with supergroup target. We
showed that for a massless particle on a supergroup with
reparameterization invariant action, the quadratic Casimir operator
becomes diagonalizable in cohomology.

Secondly, to analyze further how the on-shell spectrum of string
theory in $AdS$ backgrounds with RR flux unitarizes in conformal
gauge, we studied the stringy physical state space cohomology for a
particle on the supergroup $PSU(1,1|2)$. By keeping space-time
supersymmetry manifest at all stages, we were able to efficiently
compute the Kaluza-Klein supergravity spectrum (corresponding to the
particle limit), and to understand algebraically how unitary
superconformal multiplets arise in cohomology.

We believe our kinematical analysis shows that we should make an
effort to isolate those properties of the logarithmic conformal field
theories arising on supergroups and their cosets that will survive in
the physical state space of string theory. {From} our study it is
clear that a lot of the intricate properties of correlation functions
associated to the logarithmicity of the conformal field theories will
not survive in the BRST cohomology, simply because the states involved
in those intricate correlators are not physical. It is an important
open problem to thoroughly understand how to efficiently isolate the
stringy data within these logarithmic conformal field theories.

As a byproduct of our analysis of these questions, we showed that by
using super algebra we can very efficiently compute Kaluza-Klein
supergravity spectra on maximal bosonic subgroups of supergroups. Our
technique generalizes to cosets of supergroups, like $AdS_5 \times
S^5$ or $AdS_2 \times S^2$, etcetera, and is likely to provide a very
efficient calculation of the full Kaluza-Klein spectrum.

When working in a manifestly supersymmetric
formalism we're required to adopt super algebra representations that are
considerably larger and more intricate than Kac modules.

\section*{Acknowledgements}
We would like to thank Raphael Benichou, Denis Bernard  and Thomas Quella for 
useful discussions and constructive comments.  
Our work was supported in part by the grant
ANR-09-BLAN-0157-02 and by a PEPS grant PEPS-Physique Th\'eorique et ses
Interfaces.

\appendix

\section{A few results in  $gl(2|2)$ representation theory}
\label{gl22}
In the bulk of the paper, we refer to some results in the
representation theory of the superalgebra $gl(2|2)$ of four-by-four
super matrices.  The superalgebra $sl(2|2)$ is a subalgebra of $gl(2|2)$,
consisting of matrices of zero supertrace,
and the superalgebra $psl(2|2)$ is an ideal of $sl(2|2)$ where we mod out by the
identity matrix.
As a consequence, a representation of $psl(2|2)$ lifted to a representation
of $gl(2|2)$ will have a trivial representation of the identity matrix,
while the representation of the other extra $u(1)$ is not determined
uniquely. Note that under the extra anti-diagonal $u(1)$, the fermionic entries
of the supermatrix, corresponding to the fermionic generators,
are charged. The point of embedding the $psl(2|2)$
representations in $gl(2|2)$ representations is that we have this
extra $u(1)$ charge grading at our disposal in order to distinguish
various representation spaces.

\subsection*{Lifting representations}
The super algebra  $gl(2|2)$ has a Cartan subalgebra $\mathfrak{h}$ of diagonal
matrices $H=\mbox{diag}(h_1,h_2,h_3,h_4)$. We define linear functionals $\epsilon_i(H)=h_i$
and $\delta_j(H)=h_{2+j}$ for $ i \in \{ 1,2 \}$.
The algebra has the roots $\epsilon_i - \epsilon_j,
\delta_i-\delta_j$ for $ i \neq j$  and $ \epsilon_i - \delta_j, \delta_i-\epsilon_j$ for
$i,j \in \{ 1,2 \}$.
Here we follow \cite{Drouot} closely, and denote the weights $\lambda$ of a
$gl(2|2)$ representation by $\lambda=(a_1, a_2| b_1,b_2)$ for the weight $\lambda
= a_1 \epsilon_1 + a_2 \epsilon_2 + b_1 \delta_1 + b_2 \delta_2$.  We
have  that the weight $\lambda = (a_1,a_2|b_1,b_2)$ is atypical when one
of the numbers $a_1+b_1+1,a_1+b_2,a_2+a_1,a_2+b_2-1$ is zero. It is maximally
atypical when the weight is of the form $\lambda = (a_1,a_2|-a_2,-a_1)$.

We want to lift representations of the super algebra $psl(2|2)$ to representations
of  the super algebra $gl(2|2)$. To that end, we demand first of all that
 the identity matrix in $gl(2|2)$ be represented trivially, namely that the
 coefficients of the weight $\lambda$ satisfy 
$\sum_{i=1}^2 (a_i + b_i) =0$. It should also be clear that the spins  $j_1,j_2$ of the $sl(2) \oplus sl(2)$ 
subalgebras of both $gl(2|2)$ and $psl(2|2)$
are associated to the coefficients of the  weights $\epsilon_1 - \epsilon_2 $ and 
$\delta_1 - \delta_2$ in the weight $\lambda$ while there is also another overall anti-diagonal $u(1)$ 
associated to the coefficient of the weight $\sum_i (\epsilon_i - \delta_i)$ in the weight $\lambda$.
Therefore, a possible choice of lift of a $psl(2|2)$ representation characterized by spins $j_{1,2}$ 
is to take the weight of the lifted representation of $gl(2|2)$ 
to be $\lambda=j_1 (\epsilon_1-\epsilon_2) + 
j_2 (\delta_1-\delta_2) =(j_1,-j_1|j_2,-j_2)$. 
If we consider positive spins only, we have an atypical weight when $j_1=j_2$. 
Indeed, the Kac module built on a ground state with spins $j_1=j_2$ is atypical.
When the spins are equal, we automatically have maximal atypicality from the perspective of the algebra $gl(2|2)$.

\subsection*{The Kac composition series for maximally atypical modules}
We concentrate on the
relevant case of the atypical $psl(2|2)$ modules, which lift to
maximally atypical modules of the algebra $gl(2|2)$. Moreover, we will
focus on spins $j_1=j_2=j$ which are not too small, to avoid
exceptional cases.  We then have from the results of (\cite{Drouot} theorem 4.1.5),
that the Kac modules that appear in the Kac
decomposition series of the projective cover are
the modules $K(j,-j|j,-j)$,
$K(j,-j+1|j-1,-j)$, $K(j+1,-j|j,-j-1)$ as well as
$K(j+1,-j+1|j-1,-j-1)$.
When we restrict to the $psl(2|2)$ action, these Kac modules of $gl(2|2)$ 
correspond to $psl(2|2)$ Kac modules $K(j,j)$, $K(j-1/2,j-1/2)$,
$K(j+1/2,j+1/2)$ and $K(j,j)$. Their anti-diagonal $U(1)$ charges
(divided by two) distinguish the first and last $K(j,j)$
representations.  Their anti-diagonal gradings are $0,1,1,2$ respectively. 
\subsection*{The composition series}
We can also borrow the result for the composition series of the
projective cover in terms of short multiplets from the $gl(2|2)$
result. Indeed, the result of (\cite{Drouot} corollary 4.1.5 and lemma 4.1.6)
is used in figure 7 representing the composition series of the
projective representation in terms of irreducible modules drawn in the
bulk of the paper (and \cite{Drouot} contains even more detail).  Thus, through the
embedding, we gained that we are able to distinguish short representations by their
anti-diagonal $u(1)$ charge, and that we can borrow freely from $gl(2|2)$ representation 
theory where we can apply the Berenstein-Gelfand-Gelfand duality theorem.

\end{document}